\documentstyle[psfig]{elsart}

\begin{document} 

\begin{frontmatter}

\title{Observing the Ultrahigh Energy Universe with {\it OWL} Eyes}

\author{F.W. Stecker$^1$, J.F. Krizmanic$^1$,L.M. Barbier$^1$, E. Loh$^2$, }
\author{J.W. Mitchell$^1$, P. Sokolsky$^3$, and R.E. Streitmatter$^1$}

\address{$^1$Laboratory for High Energy Astrophysics,
NASA Goddard Space Flight Center, Greenbelt, Maryland 20771, USA\\
$^2$ National Science Foundation,
Arlington, Virginia 22230, USA\\
$^3$ High Energy Astrophysics Institute, University of Utah\\
Salt Lake City, Utah 84112, USA}

\begin{abstract}               
The goal of the Orbiting Wide-field Light-collectors ({\it OWL}) mission is to study the origin and
physics of the highest energy particles known in nature, the
ultrahigh energy cosmic rays (UHECRs). The {\it OWL} mission consists of  
telescopes with UV sensitive cameras on two satellites
operating in tandem to view in stereo the development of the giant particle showers induced in the Earth's
atmosphere by UHECRs. This paper discusses the characteristics of the {\it OWL} mission.
\end{abstract}




\end{frontmatter}

\section{The {\it OWL} Satellite Detectors}

The {\it OWL} (Orbiting Wide-field Light-collectors) mission is designed to obtain data on ultrahigh energy cosmic rays (UHECR) 
and neutrinos in order to tackle the fundamental 
problems associated with their origin \cite{Stecker}. 
The {\it OWL} mission is designed to provide the event statistics and
extended energy range that are crucial to addressing these issues. To accomplish this,
{\it OWL} makes use of the Earth's atmosphere as a huge ``calorimeter"
to make stereoscopic measurements
of the atmospheric UV fluorescence produced by air shower particles. This is the most accurate
technique that has been developed for measuring the energy, arrival direction, and
interaction characteristics of UHECR {\cite{Streitmatter}}. 
To this end, {\it OWL} will consist of a pair of satellites placed in tandem in a low inclination, medium altitude orbit. The {\it OWL} telescopes will point down 
at the Earth and will together point at a section of atmosphere about the size of the state of Texas ($\sim 6 \times 10^5$ km$^2$), obtaining a much greater sensitivity than present ground based 
detectors. The ability of of {\it OWL} to detect cosmic rays, in units of km$^2$ sr, is called the aperture. The instantaneous aperture at the highest energies is $\sim 2 \times 10^6$ km$^2$ sr (see Sect. 2.3). The effective aperture, reduced by the effects of the moon, man-made light, and clouds, will be conservatively $\sim 0.9 \times 10^5$ km$^2$ sr. 
For each year of operation, {\it OWL} will have 90 times the aperture of the ground based {\it HiRes} detector and 13 times the aperture of the {\it Pierre Auger} detector
array (130 times its most sensitive 
``hybrid" mode). The {\it OWL} detectors will observe the UV flourescence light from the giant air showers produced by UHECR on the dark side of the Earth. 
They will thus produce a stereoscopic picture of the temporal and spatial development of the showers. 

{\it OWL} has been the subject of extensive technical studies, examining all aspects of the 
instrument and mission. This paper gives an overview of {\it OWL} as based on these studies. The technical details, as well as discussion of the science, including
ultrahigh energy neutrino science with {\it OWL}, 
can be found at {\tt http://owl.gsfc.nasa.gov}. The {\it OWL} baseline instrument and mission can be realized using current technology.
The baseline {\it OWL} instrument, shown in Figure 1, is a large f/1 Schmidt camera with a $45^{\circ}$ full field-of-view (FOV) and a 3.0 meter entrance aperture. The entrance 
aperture is filled with a Schmidt corrector. The deployable primary mirror has a 7 m diameter. The focal plane has an area of 4 m$^2$ segmented into approximately 
500,000 pixels distributed over 1300 multianode photomultiplier tubes. Each pixel is read out by an individual electronics chain and can resolve single photoelectrons. 
Taking obscuration by the focal plane and by the members supporting the focal plane and corrector plate into account, the effective aperture of the instrument is about 
3.4 m$^2$. A deployable light shield, not shown in Figure 1, covers the instrument and a redundant shutter is used to close off the aperture during non-observing periods. 
A UV laser is located at the back of the focal plane and fires through the center of the corrector plate to a small steering mirror system. 
Laser light reflected by clouds is detected and measured using the {\it OWL} focal plane. {\it OWL} is normally operated in stereo mode and the instruments view a common volume of atmosphere. However, the instruments are independent and the focal plane has been designed for a time resolution of 0.1 $\mu$s so that monocular operation can be supported (with reduced performance) if one instrument fails.
The instrument weight is $\sim 1800$ kg and total power consumption is $\sim 600$ W. The amount of data generated by the instrument is determined by calibration and atmospheric monitoring and averages 150 kbps over any 24 hour period.

\begin{figure}[h]
\begin{center}
\mbox{\psfig{figure=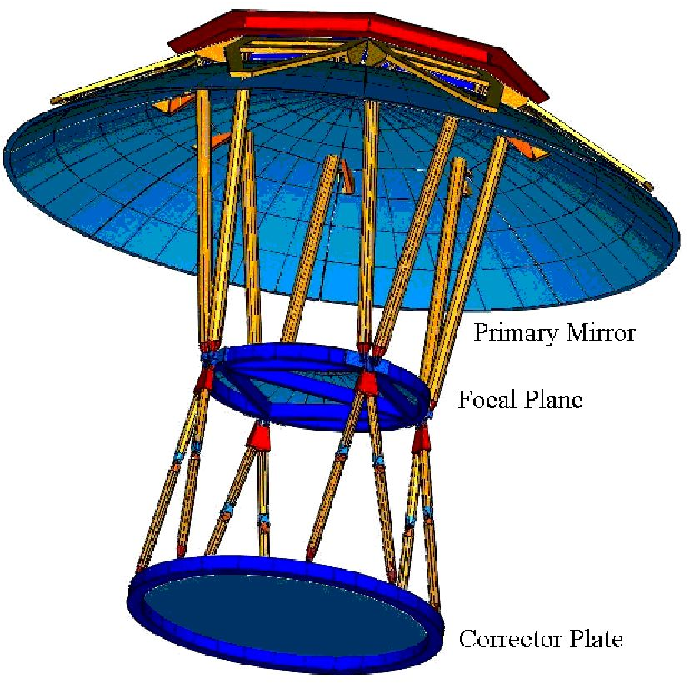,height=9cm}}
\hspace{1.cm}
\mbox{\psfig{figure=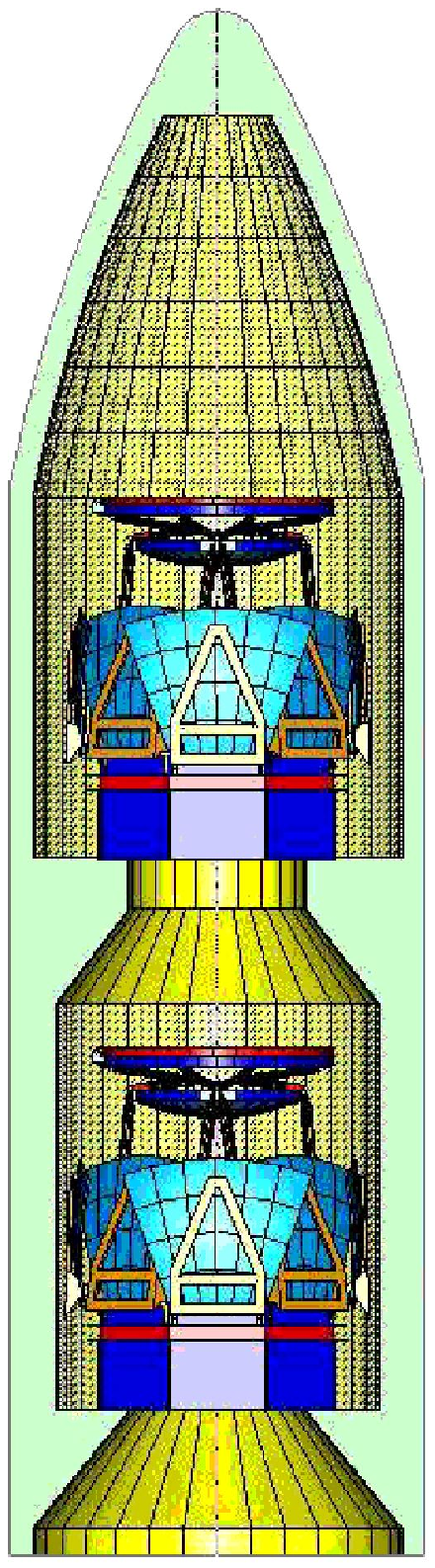,height=10cm}}
\end{center}

\caption{Left Figure: Schematic of the Schmidt optics that form an {\it OWL} ``eye'' in the deployed configuration.  The spacecraft bus, light shield, and shutter are not shown.  Right Figure: Schematic of the stowed {\it OWL} satellites in the launch vehicle.}
\end{figure}

The satellites are launched together on a Delta rocket into a 1000 km circular orbit with a nominal inclination of $10^{\circ}$. Figure 1 shows both satellites stowed for launch.
Following launch, the two satellites will fly in formation with a separation of 10 to 20 km for about 3 months to search for
upward going showers from $\nu_{\tau}$'s propagating thorugh the Earth. The spacecraft will then separate to 600 km for $\sim$ 2.5 years to measure the high-energy end of the 
UHECR spectrum. Following this period, the altitude is reduced to 600 km and the separation to 500 km in order to measure the cosmic ray flux closer to $10^{19}$ eV. 
The choice of two different orbits is the result of a tradeoff among the collecting power, the FOV of the optical system, and the energy threshold of the instrument. 
The {\it OWL} instruments function independently and data aquisition does not require space-to-space communication. Events are triggered separately in each instrument using
hierarchical hardware and software algorithms to suppress background and localize shower tracks on the focal plane. A hardware trigger, designed for very high efficiency and 
minimal dead time, initiates data acquisition in a limited region of the focal plane around the point where the trigger was generated. All shower data acquired will be telemetered to the ground. Data from both instruments are combined on the ground using GPS time stamps.
After an event location and crude track direction are determined by the trigger system, a series of laser shots are taken along the track by each instrument to determine 
local atmospheric conditions in the region of triggered events. In addition, a random scan of the FOV is made at approximately one laser shot per second in order to 
determine the average cloud obscuration. These measurements are complemented by data obtained from geostationary and polar orbiting IR imaging satellites.

\begin{figure}


Figure 2: Two OWL satellites in low-Earth orbit observing the flourescent
track of a giant air shower. The shaded cones illustrate the field-of-view
for each satellite.


PS FILE OF FIGURE TOO LARGE FOR ASTRO-PH

\end{figure}

\section{UHECR Calorimetry by UV Fluorescence}

With the fluorescence technique, a fast, highly pixelized camera (or ``eye") is used to resolve
both the spatial and temporal development of the shower.
This detailed information provides a powerful tool for determining the nature of the
primary particle. The UV emission, principally in the 300 to 400 nm range, is 
isotropic and the
camera can view the shower from any direction, except almost
directly toward the camera. In the exceptional case, the camera 
may still be utilized as a Cherenkov detector. Thus, a single camera can view a
particles incident on the Earth from a hemisphere of sky. 
Each camera images the projection
of the shower onto a plane normal to the viewing direction. 

In monocular operation, precision
measurements of the arrival times of UV photons from different parts of the shower track must
be used to partly resolve spatial ambiguities. The angle of the shower relative to the viewing plane
is resolvable using differential timing. Resolving distance, however, requires that the pixel
crossing time be measured to an accuracy that is virtually impossible to achieve in a real instrument
at orbit altitudes.
Stereoscopic observation resolves both of these ambiguities. In stereo, fast
timing provides supplementary information to reduce systematics and improve the resolution
of the arrival direction of the UHECR. By using stereo, 
differences in atmospheric absorption or scattering of the UV light can be determined. The results
obtained by the {\it HiRes} collaboration viewing the same shower in both modes have clearly demonstrated
the desirability of stereo viewing. 

\subsection{Instrument Details}

Even at large zenith angles from the {\it OWL} orbits, the maximum extent of the showers is several tens of kilometers. Thus, measuring the longitudinal profile of the cascade leads to a natural scale of $\sim$ 1 km. The corresponding optical angular resolution required depends upon the orbit altitude; observation from 1000 km implies an angular resolution of 1 mrad.
The optical system is a low resolution imager with more similarities to a microwave system than to a precision optical telescope. This allows the optical design to be simplified and has resulted in the selection of a simple, wide-angle ($45^{\circ}$ FOV), Schmidt camera design as shown in Figure 1. The Schmidt corrector has a spherical front surface and an aspheric back surface, while the primary mirror has a slight aspheric figure. The focal plane is a spherical surface tiled with flat detector elements.
The corrector is slightly domed for strength. The primary is made of lightweight composite material with a central octagonal section and eight petals that fold upward for launch. The entire optical system is covered by an inflatable light and micrometeroid shield and is closed out by a redundant shutter system. The shield is composed of a multi-layer material with kevlar layers for strength.

The focal plane detector system has a total area of 4 m$^2$ divided into $5.4 \times 10^5$ pixels, each with an area of 7.4 mm$^2$ and able to detect ultraviolet light at the single photoelectron level.  The dead area between pixels or groups of pixels is minimized both to maximize the detected signal and to insure that most showers will produce contiguous tracks. The detector and readout electronics measure incident light in 0.1 $\mu$s intervals to track the shower as it crosses the field of view of the pixel (3.3 $\mu$s for a shower perpendicular to the viewing direction to cross 1 km). The timing information helps improve reconstruction systematics and angular resolution and supports potential monocular operation. The focal plane incorporates an absorption filter with a tailored bandpass between 330 and 400 nm to suppress optical background.

\subsection{Atmosphere and Cloud Monitoring}

Events observed by {\it OWL} will occur over an observing area which moves across the globe at a speed of 7 km s$^{-1}$. This moving area will cover regions with  variable amounts of clouds with altitudes from sea level to 15 km and variable boundary layer aerosols. Most of the giant air showers will lie below 10 km. Consistency between the stereo views of an event provides a powerful tool for understanding whether the profiles of individual events have been altered by scattering through intervening high clouds or aerosol layers. In addition, {\it OWL} will use a steerable UV laser beam to scan the region of the event as a simple altimeter for cloud heights to provide real-time characterization of the atmosphere. To this end, {\it OWL} will employ a diode-pumped Nd:YAG laser with about 75 mJ output at 1064 nm. Third-harmonic generation is used to obtain a 355 nm beam with an energy of about 15 mJ, a pulse duration of 5 ns, and an emittance of about 1 mrad. The laser fires into a small steerable mirror system capable of slewing to any point in the {\it OWL} FOV in less than one second. During the {\it OWL} mission life, it is expected that the laser will fire on the order of $10^9$ times. While it is impractical to use the {\it OWL} laser to fully map the atmosphere, it will be used to provide a sparse scan of the full FOV. Accumulated over many viewing passes, this information will provide an excellent statistical basis for understanding the aperture. This information will be complemented by geostationary and polar-orbiting IR satellite data to characterize the fraction of clear pixels in the detector aperture.

\subsection{{\it OWL} Monte Carlo Simulations}

Monte Carlo simulations of the physics and response of orbiting instruments to the UV air fluorescence signals are crucial to the development of {\it OWL}. One such Monte Carlo has been developed at the NASA Goddard Space Flight Center {\cite{Krizmanic}}. The simulation employs a hadronic event generator that includes effects due to fluctuation in the shower starting point and shower development, charged pion decay, neutral pion reinteraction, and the LPM effect. Because the {\it OWL} baseline imaging requirements are rather insensitive to lateral shower size, the hadronic generator creates individual 1-dimensional shower parameterizations characterized as a 4-parameter Gaisser-Hillas function. Each air shower is developed in a sequence of fixed time intervals of 1 $\mu$s and the resulting charged particles are used to generate air fluorescence and Cherenkov signals. The fluorescence signal is corrected for the pressure and temperature dependence of the atmosphere and large-angle scattering of the Cherenkov light into the viewing aperture of the instrument is accounted for. Once the UV light signal is generated, it is propagated out of the atmosphere to the orbit altitude including light losses by Rayleigh scattering and ozone absorption. Response functions for the optical transmission and focal plane spot size as functions of shower viewing angle are based on the results of optical ray tracing modeling. The UV signal is attenuated by the filter response, mapped onto the focal plane array, and convolved with the wavelength response of the photocathode. The resultant pixel signals are Poisson fluctuated to obtain a photoelectron signal in each pixel for each time step. At the peak of a $10^{20}$ eV shower, the typical signal obtained in a single pixel crossing time (3.3 $\mu$s) is $\sim$ 10  photoelectrons and depends on the location of the shower in the FOV. The background in the same time interval is about 0.6 photoelectrons. For $10^{20}$ eV protons, event reconstruction results show an energy resolution of about 16\%
and good resolution of the shower maximum. The Monte Carlo simulations have indicated agreement to within $1^{\circ}$ in each view between generated shower tracks and tracks reconstructed using a moment of inertia method based on the amplitude measurements. This agreement may be improved by incorporating timing information.

\begin{figure}
\centerline{\psfig{figure=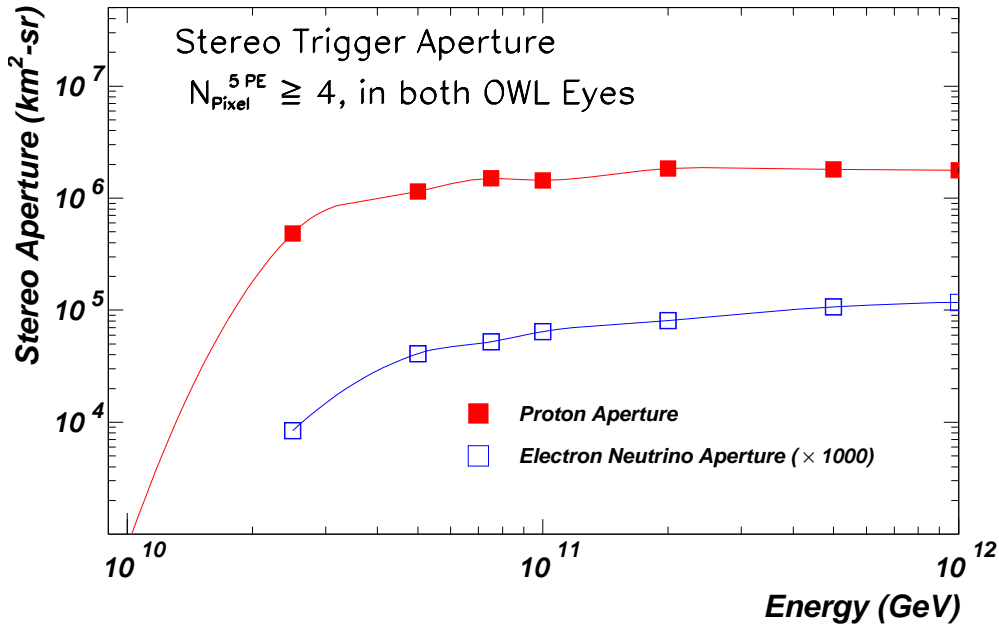,height=10cm}}
\caption{Instantaneous aperture for proton-induced and deep $\nu_{e}$-induced giant air-showers as a function of energy.}
\end{figure}

The number of events detected by OWL for a monoenergetic isotropic flux of protons and $\nu_{e}$'s with a standard model cross section is calculated 
by the Monte Carlo program, yielding the detection aperture 
as a function of energy, simulated trigger, and orbit parameters. Figure 3 shows the resultant proton and neutrino aperture for an altitude of 1000 km and a separation of 500 km. The asymptotic instantaneous proton aperture is $\sim 2 \times 10^6$ km$^2$sr. The $\nu_{e}$ aperture determination includes the requirement that the observed starting point of the air shower, $X_{start} \ge 1500$ g cm$^{-2}$ in slant depth. 

\subsection{Viewing Efficiency}

The man-made light and the polar aurora make a high inclination orbit undesirable. The {\it OWL} satellites will be placed in an orbit lying within 10$^{\circ}$ of the equator. Each {\it OWL} eye must remain shuttered until the viewed Earth has entered full darkness, free from sunlight or moonlight. It then opens and after $\sim 1$ min can begin observations. Requiring that both instruments only take data in full darkness results in an overall duty factor of 14.4\%.
Sunlight, moonlight, lightning, man-made light, oceanic biofluorescence, and high altitude clouds further reduce the observation efficiency of {\it OWL}. The effects of clouds are more difficult to evaluate since they may affect only a fraction of the aperture {\cite{Clouds1,Clouds2}}. Global equatorial thunderstorm activity and the incidence of clouds above 3 km altitude gives a further reduction in efficiency by about 70\% under the assumption that these preclude observation in the affected region {\cite{Clouds3}}. This gives an overall efficiency of Å 4.4\%, and an overall effective equivalent aperture of $\approx 0.9 \times 10^5$ km$^2$ sr. The 4.4\% overall efficiency is largely determined by orbit, moonlight and cloud conditions and will be approximately the same for {\it any} observation of giant air showers from space using the fluorescence technique. For showers of sufficiently high energy, the use of an adaptive threshold could improve the efficiency by accepting events that occur in partial moonlight.

\section{Conclusion}

The {\it OWL} ``eyes'', orbiting space-based telescopes with UV sensitive
cameras, will use large volumes of the Earth's atmosphere as a detecting medium to trace the atmospheric
flourescence trails of large numbers of giant air showers produced by ultrahigh energy cosmic rays and neutrinos, potentially 
to energies higher than those presently observed. Such 
missions will open up a new window, not only on
extragalactic astronomy and cosmology, but also on physics at the highest possible observed energies.

\end{document}